\documentclass[twocolumn,superscriptaddress,floatfix,preprintnumbers, nofootinbib,hyperref]{revtex4-2} 
\pdfoutput=1
\usepackage[colorlinks=true,breaklinks=true]{hyperref}
\usepackage[normalem]{ulem}
\usepackage[utf8]{inputenc}
\hypersetup{allcolors=[rgb]{0.0 0.0 0.6},linkcolor=[rgb]{0.75 0.05 0.05}}
\usepackage{amsmath,amssymb}
\usepackage{epsfig}  
\usepackage{graphicx}   
\usepackage{slashed}       
\usepackage{tikz}
\usepackage{tikz-feynman}
\usepackage{url}
\usepackage{color}
\usepackage{multirow}
\usepackage{comment}
\usepackage{amssymb}
\usepackage{makecell}
\usepackage{array}
\usepackage{booktabs}

\newcommand{\bq}{{\bf q}}

\newcommand{\beq}{\begin{equation}}
\newcommand{\eeq}{\end{equation}}

\newcommand{\portal}[1]{\makecell[c]{#1}}
\newcommand{\portalmark}[2]{\makecell[c]{#1\rlap{\ensuremath{^{#2}}}}}
\newcommand{\runinsec}[2]{%
  \par\refstepcounter{section}\phantomsection\label{#2}%
  \noindent\emph{\textbf{#1}---}\ignorespaces
}

\allowdisplaybreaks

\bibpunct{[}{]}{,}{n}{}{,}

\pgfarrowsdeclare{X}{X}
{
  \pgfutil@tempdima=0.3pt%
  \advance\pgfutil@tempdima by.25\pgflinewidth%
  \pgfutil@tempdimb=5.5\pgfutil@tempdima\advance\pgfutil@tempdimb by.5\pgflinewidth%
  \pgfarrowsleftextend{+-\pgfutil@tempdimb}
  \pgfarrowsrightextend{0pt}
}
{
  \pgfutil@tempdima=0.3pt%
  \advance\pgfutil@tempdima by.25\pgflinewidth%
  \pgfsetdash{}{+0pt}
  \pgfsetroundcap
  \pgfsetmiterjoin
  \pgfpathmoveto{\pgfqpoint{-5.5\pgfutil@tempdima}{-6\pgfutil@tempdima}}
  \pgfpathlineto{\pgfqpoint{5.5\pgfutil@tempdima}{6\pgfutil@tempdima}}
  \pgfpathmoveto{\pgfqpoint{-5.5\pgfutil@tempdima}{6\pgfutil@tempdima}}
  \pgfpathlineto{\pgfqpoint{5.5\pgfutil@tempdima}{-6\pgfutil@tempdima}}
  \pgfusepathqstroke 
}
\makeatother

\begin{document}

\title{Dark Glueball Direct Detection}

\author{Ji-Wei Li}
\email{jiweili@std.uestc.edu.cn}
\affiliation{School of Physics, The University of Electronic Science 
and Technology of China,\\
No.~2006, Xiyuan Ave, West Hi-Tech Zone, Chengdu, China}

\author{Roman Pasechnik}
\email{Corresponding: roman.pasechnik@fysik.lu.se}
\affiliation{
 Department of Physics, Lund University
 SE-223 62 Lund, Sweden
}

\author{Wei Wang}
\email{wei.wang@sjtu.edu.cn}
\affiliation{
 State Key Laboratory of Dark Matter Physics, Key Laboratory for Particle 
 Astrophysics and Cosmology (MOE),
 Shanghai Key Laboratory for Particle Physics and Cosmology,
 Shanghai Jiao Tong University, Shanghai 200240, China
}

\author{Zhi-Wei Wang}
\email{Corresponding: zhiwei.wang@uestc.edu.cn}
\affiliation{School of Physics, The University of Electronic Science and Technology of China,\\
No.~2006, Xiyuan Ave, West Hi-Tech Zone, Chengdu, China}

\smallskip

\begin{abstract}
We consider glueball dark matter (DM) in a Yang-Mills dark sector confined at $\Lambda_D$ scale and 
coupled to the Standard Model through electrically and dark-color charged vector-like fermion 
portals, with the mass scale $m_\psi$. In a simple case with two lightest mass-degenerate vector-like fermions 
with opposite electric charges the effective amplitudes with 
one $C$-odd glueball (oddball) and odd number of photons vanish, rendering the lightest $C$-odd 
spin-1 state with mass $m_\chi$ a viable DM candidate provided 
that $m_\psi\gtrsim 5.5 \Lambda_D$. 
We develop a controlled effective field theory framework with non-perturbative information supported
by QCD phenomenology leading to a quantitative prediction for coherent elastic glueball scattering 
off nuclei. We find a steep scaling of the spin-independent cross section $\sigma_{\rm SI}\propto \Lambda_D^{2.15} m_\psi^{-8}$. This implies that the sensitivity of the current and next-generation 
xenon experiments in the range of $\sigma_{\rm SI} \sim 10^{-46} - 10^{-48}$ cm$^2$ corresponds 
to $m_\psi \simeq 3-30$ GeV, respectively, for $\Lambda_D\simeq 0.55-5.5$ GeV. 
We provide a minimal UV completion of the portal sector compatible with collider phenomenology. 
Our results pave a quantitative foundation for testing glueball DM in direct-detection 
experiments.
\end{abstract}

\maketitle

\runinsec{I: Introduction.}{sec:introduction}
The particle nature of dark matter (DM) remains unknown, and increasingly stringent bounds on weakly coupled candidates have strengthened the case for a richer hidden sector. 
Among many DM models, confining dark non-Abelian sectors provide a broadly motivated 
UV-complete framework for strongly interacting DM. As in QCD of the Standard Model (SM), confinement generates 
mass gap and naturally yields composite states -- dark hadrons -- governed by strong dynamics as reviewed in Ref.~\cite{SC001}. Such sectors can also feature sizable DM self-interactions that may affect small-scale 
structure while remaining compatible with current cosmological and astrophysical constraints \cite{SC007,SC002}.

A systematic framework for secluded strong dynamics is provided by the ``hidden valley'' paradigm
\cite{SC011,SC010,SC013}, which emphasizes that portals to a confining sector can yield complex hidden
hadronization and, often, long-lived states with displaced decays \cite{SC012}. In the minimal confining 
formulation, i.e.~a pure Yang-Mills theory, the infrared spectrum consists of glueballs.
Pure-glue hidden valleys and their portal-induced decays were systematically explored in
Refs.~\cite{SC014,SC015}, while the necessary nonperturbative inputs (glueball masses and 
matrix elements) are anchored by lattice determinations of the glueball spectrum \cite{SC016,SC017}.
This ``dark glueball'' limit is thus theoretically clean yet phenomenologically rich: 
it is strongly coupled in the infrared, but admits controlled effective descriptions 
once matched to nonperturbative information.

Dark glueballs have long been explored as DM candidates. Early studies of hidden $\mathrm{SU}(N)$ 
glueball DM \cite{SC022,SC025} established the cosmological picture, emphasizing number-changing 
reactions and dark-sector thermodynamics. In ultraviolet-motivated formulations, glueballs can be 
overproduced or decay too late (the ``dark glueball problem'' \cite{SC026}), while nonstandard 
cosmologies can substantially reshape the viable parameter space \cite{SC024}. Lattice-calibrated thermal EFTs further show that relic-density predictions are naturally tied to the confinement scale and the dark-to-visible temperature ratio, and can deviate significantly from conventional estimates \cite{SC064,SC065}, motivating nonperturbative inputs throughout glueball phenomenology.

Confining dark sectors are amenable to multi-messenger probes. A first-order confinement transition 
may generate a stochastic gravitational-wave (GW) background \cite{SC061,SC060}. Using lattice-informed 
effective descriptions, dark $\mathrm{SU}(N)$ Yang-Mills confinement and its implications for GW signals 
have been studied in \cite{SC062,SC063,Kang:2021epo,SC066}. Complementary sensitivity comes from 
indirect detection when DM annihilates into dark glueballs that subsequently decay to SM 
states \cite{SC058}, and from collider searches for exotic Higgs decays and long-lived signatures 
of dark confinement \cite{SC057,SC059}. Together, these channels motivate a unified program 
in which cosmology, astrophysics, and laboratory searches jointly probe strong hidden dynamics.

Direct detection remains the least developed part of the glueball-DM program: unlike pointlike DM, 
minimal portals induce high-dimensional couplings, and recoil rates require controlled matching 
from ultraviolet portals to nonperturbative glueball amplitudes. Photon-mediated scattering provides 
a generic route for neutral but polarizable composites \cite{SC052,SC056}, while heavy-mediator portals 
can be organized within gluon/chromo-Rayleigh EFTs \cite{SC053,SC054}. These encode the DM response 
in local operators, whereas for glueballs the scattering amplitude must be supplied nonperturbatively, 
and related dynamics can yield extremely suppressed electromagnetic couplings in other composite 
settings \cite{SC067}. A key open question is whether glueball DM can be placed on a comparably 
quantitative footing for \emph{direct} searches, with controlled matching between ultraviolet 
portals, nonperturbative glueball matrix elements, and experimentally relevant scattering amplitudes.

In this Letter, we address this question by adopting nonperturbative information from QCD phenomenology
and establish a tensor-Pomeron-inspired EFT framework for direct detection. For electrically charged portal 
fermions with $m_\psi\!\gg\!\Lambda_D$ and $C$-odd glueballs below $2m_\psi$, integrating out the portal 
yields two-photon operators that control nuclear scattering via two off-shell (Coulomb) photons.
We study two minimal benchmarks: a single charged fermion, and a two-flavor vector-like portal with 
approximate global $\mathrm{SU}(2)_D$ and opposite electric charges, which suppresses odd-photon couplings 
and render lightest ``oddball'' glueballs effectively stable. By linking glueball lifetimes to direct-detection 
rates within a controlled EFT framework, our results pave a quantitative foundation for direct 
searches of glueball DM.

\runinsec{II: Light glueballs and fermionic portal.}{sec:portal}
In this work, we consider a generic dark $\mathrm{SU}(N)$ gauge theory confining at a scale $\Lambda_D$. 
Its low-energy spectrum consists of glueballs denoted as $\chi$, i.e.~purely gluonic bound states characterized by their 
spin and discrete quantum numbers $J^{PC}$. We focus on the lowest-lying $J=0,1$ states summarized 
in Table~\ref{glueball_classification}, together with their $P,C$ assignments and characteristic masses.

To connect the dark glueball sector to the SM, we introduce electrically charged portal 
fermions with mass scale $m_\psi$ and work in the hierarchical regime $m_\psi \gg \Lambda_D$, such 
that the lightest composite states remain glueballs (glueball states of interest lie below the portal-pair 
threshold $2m_\psi$). For most of this Letter we adopt a low-energy EFT viewpoint -- dark Yang-Mills coupled 
to QED through loops of heavy charged fermions -- and defer a concrete electroweak (EW) completion and collider 
constraints to Sec.~\ref{sec:UV}. We consider two minimal benchmark portal structures: (I) a single electrically 
charged Dirac fermion $\psi$, and (II) two mass-degenerate electrically charged Dirac fermions $\psi_+$ and $\psi_-$
forming a doublet of an (approximately) conserved global $\mathrm{SU}(2)_D$ dark-flavor symmetry, with opposite 
electric charges $Q_{\psi_+}=-Q_{\psi_-}$. 

\paragraph{Scenario I (heavy portal).}
For a minimal single-$\psi$ portal with sufficiently large $m_\psi \gtrsim 10~\mathrm{TeV}$ required by
astrophysical bounds on decaying DM \cite{Blanco:2018esa,Munbodh:2024ast}, DM would be dominated by the lightest $C$-even scalar $0^{++}$ 
and pseudoscalar $0^{-+}$ glueballs. In this regime, the portal-induced photon couplings mediated by 
fermion-box loop are extremely suppressed, leading to a negligible elastic glueball-nucleus cross section, 
below $10^{-70}\,\mathrm{cm}^2$.

\paragraph{Scenario II (light portal).}
In the approximate $\mathrm{SU}(2)_D$ doublet portal with vector-like $m_{\psi_+}\simeq m_{\psi_-}\equiv m_\psi$ 
and $Q_{\psi_+}=-Q_{\psi_-}$, the lightest $1^{+-}$ and next-to-lightest $1^{--}$ $C$-odd vector glueballs (oddballs) 
are (nearly) stable being viable DM candidates while heavier oddballs are expected to be Boltzmann suppressed.
In the dark-flavor symmetry limit, any loop amplitude involving a single oddball and $2n+1$ photons vanishes due to 
$Q_{\psi_+}^{2n+1}+Q_{\psi_-}^{2n+1}=0$, forbidding radiative cascades such as $1^{+-},1^{--}\to 0^{++}+\gamma$.
The elastic glueball-nucleus scattering would be dominated by the two-photon exchange which will be studied 
in detail below. For light portals, it is imperative to verify their consistency with collider 
constraints (Sec.~\ref{sec:UV}). Provided that $1^{+-}$ is heavier than $0^{++}$, its relative abundance 
would be Boltzmann-suppressed, thus, mitigating its overproduction. The portal thermal history differs in 
the two scenarios mainly because of their large mass hierarchy. While in Scenario~II the fermions annihilate 
away before confinement, in Scenario~I a small residual $\psi\bar\psi$ population may survive down to 
$T\sim\Lambda_D$ but the corresponding bounds states are short-lived.

Table~\ref{glueball_classification} summarizes the lowest $J=0,1$ glueballs and their leading portal-induced
couplings to photons scaling with $\alpha_D,\alpha,\Lambda_D$ and $m_\psi$ and 
indicates which benchmark controls the late-time phenomenology. It also flags that the $C$-even vectors 
$1^{++}$ and $1^{-+}$ can be long-lived since the Landau-Yang theorem forbids $1^{++,-+}\!\to\!\gamma\gamma$, 
whereas in Scenario~II the dominant decay mode $1^{++,-+}\!\to\!\gamma(\gamma^*\!\to\!\ell^+\ell^-)$ 
is forbidden below the threshold of $2m_e$.
%%%%%%%%%%%%%%%%%%%%%%%%%%%%%%%%%%%%%%%%%%%%%%%%%%%%%%%
\begin{table}[t]
\centering
\squeezetable
\renewcommand{\arraystretch}{1.3}
\setlength{\tabcolsep}{5pt}
\footnotesize

\begin{ruledtabular}
\begin{tabular}{c|c|c|c|c}
\multicolumn{1}{c}{$J^{PC}$} &
\multicolumn{1}{c}{Mass/$m_{0^{++}}$} &
\multicolumn{1}{c}{Decay operator} &
\multicolumn{1}{c}{Scaling} &
\multicolumn{1}{c}{Portal} \\
\hline
% ---------------- J = 0 ----------------
$\mathbf{0}^{++}$  & $1$      & $G^{2}F^{2}$
& \multirow{2}{*}{$\alpha_D^2\alpha^2\,\Lambda_D^{9}m_{\psi}^{-8}$}
& \multirow{2}{*}{\portal{I}} \\
\cline{1-3}
$\mathbf{0}^{-+}$  & $\sim1.55$ & $G\tilde{G}F\tilde{F}$
&  &  \\
\hline

% ---- 0^{+-} and 0^{--} share Portal = II across 4 rows
\multirow{2}{*}{$\mathbf{0}^{+-}$} & \multirow{2}{*}{$\sim2.7$}
& $G^{3}F$
& $\alpha_D^{3}\alpha\,\Lambda_D^{9}m_{\psi}^{-8}$
& \multirow{4}{*}{\portal{I}/\portal{II}} \\
\cline{3-4}
&  & $G^{3}F^{3}$
& $\alpha_D^{3}\alpha^{3}\,\Lambda_D^{17}m_{\psi}^{-16}$
&  \\
\cline{1-4}
\multirow{2}{*}{$\mathbf{0}^{--}$} & \multirow{2}{*}{\textemdash}
& $\tilde{G}G^{2}F$
& $\alpha_D^{3}\alpha\,\Lambda_D^{9}m_{\psi}^{-8}$
&  \\
\cline{3-4}
&  & $\tilde{G}G^{2}\tilde{F}F^{2}$
& $\alpha_D^{3}\alpha^{3}\,\Lambda_D^{17}m_{\psi}^{-16}$
&  \\
\hline
% ---------------- J = 1 ----------------
$\mathbf{1}^{++}$ & \textemdash
& \multirow{2}{*}{$G^{3}F$}
& \multirow{2}{*}{$\alpha_D^{3}\alpha\,\Lambda_D^{9}m_{\psi}^{-8}$}
&  \multirow{2}{*}{\portal{I}/\portalmark{II}{a}} \\
\cline{1-2}  %\cline{5-5}
$\mathbf{1}^{-+}$ & $\sim2.49$
&  &  & %\portalmark{II}{a} 
\\
\hline

% ---- 1^{+-} and 1^{--} share Portal = II across 4 rows
\multirow{2}{*}{$\mathbf{1}^{+-}$} & \multirow{2}{*}{$\sim1.78$}
& $G^{3}F$
& $\alpha_D^{3}\alpha\,\Lambda_D^{9}m_{\psi}^{-8}$
& \multirow{4}{*}{\portal{I}/\portal{II}} \\
\cline{3-4}
&  & $G^{3}F^{3}$
& $\alpha_D^{3}\alpha^{3}\,\Lambda_D^{17}m_{\psi}^{-16}$
&  \\
\cline{1-4}
\multirow{2}{*}{$\mathbf{1}^{--}$} & \multirow{2}{*}{$\sim2.44$}
& $\tilde{G}G^{2}F$
& $\alpha_D^{3}\alpha\,\Lambda_D^{9}m_{\psi}^{-8}$
&  \\
\cline{3-4}
&  & $\tilde{G}G^{2}\tilde{F}F^{2}$
& $\alpha_D^{3}\alpha^{3}\,\Lambda_D^{17}m_{\psi}^{-16}$
&  \\
\end{tabular}
\end{ruledtabular}
\caption{
Scenario I (“heavy portal”): $m_\psi \gtrsim 10~\mathrm{TeV}$ (unfavorable for direct detection).
Scenario II (“light portal”): an approximately $SU(2)_D$-symmetric \emph{dark-flavor} doublet $(\psi_+,\psi_-)$ 
with $m_{\psi_+}\simeq m_{\psi_-}\equiv m_\psi \lesssim \mathcal{O}(100~\mathrm{GeV})$ and opposite 
electric charges $Q_{\psi_+}=-Q_{\psi_-}$. For case $^{a}$, the glueball is stable for mass below $2m_e\sim1~\mathrm{MeV}$. Here, $\alpha_D$ and $\alpha\simeq 1/137$ denote dark $SU(N)$ and fine-structure couplings.
}
\label{glueball_classification}
\end{table}
%%%%%%%%%%%%%%%%%%%%%%%%%%%%%%%%%%%%%%%%%%%%%%%%%%%%%%%

\begin{figure}
  \centering
  \includegraphics[width=\columnwidth]{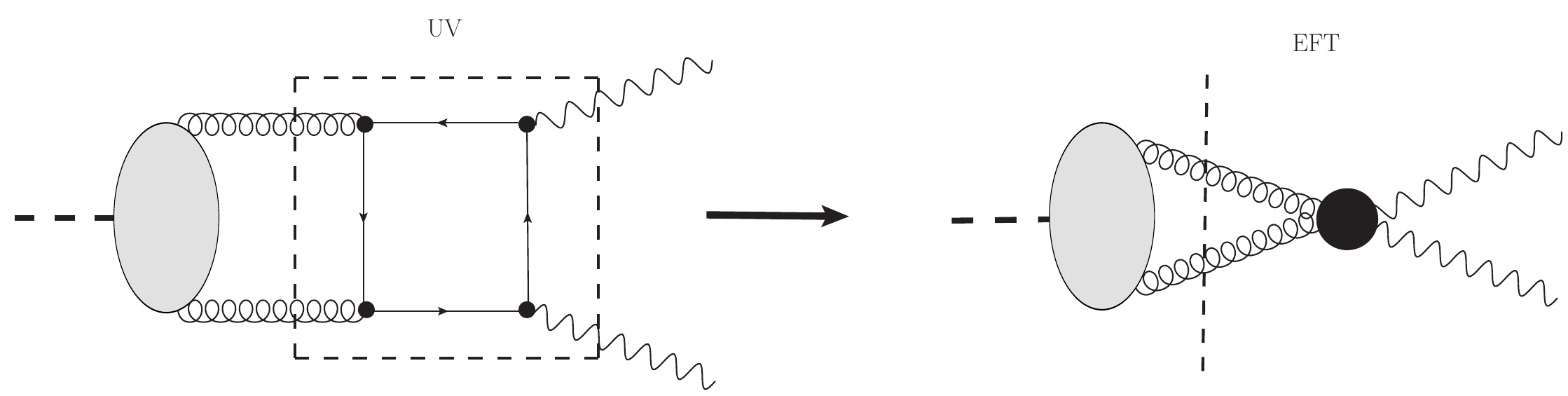}
  \caption{UV-to-EFT matching for the effective two-photon-glueball vertex. The portal-fermion 
  loop (left) is represented in the EFT as a dim-8 operator (right) that mediates 
  both $\chi\to\gamma\gamma$ and $\chi A$ elastic scattering via two off-shell photons.}
  \label{fig:decay}
\end{figure}

Transitioning to direct detection, the portal-induced EFT operators in Table~\ref{glueball_classification} 
also mediate scattering in matter. In the Coulomb regime, for either portal scenario, $\chi A\to\chi A$ proceeds 
via two off-shell photons and is governed by the doubly-virtual Compton tensor $T^\chi_{\mu\nu}$ from 
the dim-8 $F^2G^2$ matching (Fig.~\ref{fig:decay}). We now compute this two-photon 
exchange for a generic glueball state $\chi(J^{PC})$, with $J=0,1$.
%%%%%%%%%%%%%%%%%%%%%%%%%
%%%%%%%%%%%%%%%%%%%%%%%%%

\runinsec{III: Elastic glueball scattering off a nucleus.}{sec:Elastic}
Consider elastic $\chi A\to \chi A$ scattering of a dark glueball state $\chi(J^{PC})$ off a nucleus $A$ 
(mass number $A$, charge $Z$) at rest. We focus on the coherent Coulomb regime relevant for Galactic DM, 
$v_{\rm rel}\sim 10^{-3}$, where the interaction is dominated by impact parameters $b\gtrsim R_A$ 
(equivalently, $|\mathbf q|\lesssim 1/R_A$) and is encoded by the nuclear form factor ${\cal F}_A(Q^2)$. 
At leading order in $\alpha$, the process proceeds via exchange of two off-shell photons, 
Fig.~\ref{fig:elastic}. The glueball side is described by the doubly-virtual Compton 
tensor $T^{\chi}_{\mu\nu}$ for $\gamma^\ast(k_1)\chi(p)\to\gamma^\ast(k_2)\chi(p')$, generated 
by the portal-induced dim-8 operator schematically denoted $F^2G^2$ (see Table~\ref{glueball_classification}). 
In fact, $T^{\chi}_{\mu\nu}$ factorizes into a short-distance matrix element $\gamma^\ast\gamma^\ast\to g_D g_D$ 
and a long-distance one, $\langle\chi|G^2|\chi\rangle$. The full elastic amplitude $\chi(p) A(P)\to \chi(p') A(P')$ 
is written in terms of $T^{\chi}_{\mu\nu}$ as:
%%%%%%%%%%%%%%%%%%%%%%%%%%%%
\begin{figure}
  \centering
  \includegraphics[width=\columnwidth]{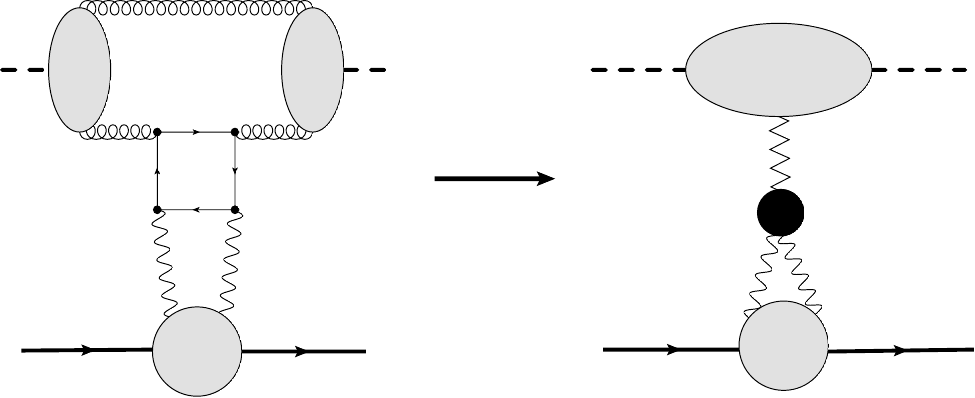}
  \caption{Elastic glueball-nucleus scattering in the Coulomb regime. 
Left panel shows two-photon exchange mediated by portal-box, while right one illustrates the
factorized form with the glueball Compton tensor $T^\chi_{\mu\nu}$, modeled as $T^\chi_{\mu\nu}\sim \Gamma^{(\gamma\gamma\mathbb{P}_D)}\,\Delta\,\Gamma^{(\chi\chi\mathbb{P}_D)}$ 
in the dark-Pomeron framework.}
  \label{fig:elastic}
\end{figure}
%%%%%%%%%%%%%%%%%%%%%%%%%%%%

\begin{equation}
\label{eq:MBornCov-Arest-upd}
\mathcal M_\chi = (4\pi\alpha)^2 Z^2\int\frac{d^4k}{(2\pi)^4}
\frac{J_A^{\mu}(k_1)J_A^{\nu}(k_2)T_{\mu\nu}(p,q;k)}{(k_1^2+i\epsilon)(k_2^2+i\epsilon)}\,,
\end{equation}
in terms $k\equiv (k_1-k_2)/2$ the relative photon momentum, $q\equiv p'-p$ the total exchanged momentum in the $t$-channel, with $|\mathbf q|\simeq\sqrt{-t}$ and $q^0\simeq-|\mathbf q|^2/(2m_A)\ll |\mathbf q|$, and $J_A^{\mu}$ the elastic current of a spinless nucleus (normalized to $Z$) defined as
\begin{equation}
J_A^\mu(k)\equiv \langle A(P')|J^\mu_{\rm em}(0)|A(P)\rangle
=(P{+}P')^\mu\, {\mathcal F}_A(Q^2),
\end{equation}
with $Q^2 \equiv -k^2 > 0$ and $\mathcal F_A(0)=1$. In the $A$ rest frame, 
\begin{equation}
J_A^\mu(k) \simeq (2m_A,\,\mathbf{0})\,\mathcal F_A(Q^2) \,,    
\end{equation}
so that the scattering is driven by exchanges of the longitudinally polarized virtual photons as expected in the Coulomb regime, while transverse components of $J_A^\mu$ are suppressed by $|{\bf q}|/m_A\ll 1$. Carrying out $k^0$ integral, one obtains
\begin{eqnarray}
\label{eq:M-final-exactProj}
\mathcal M_\chi &\simeq& e^4 Z^2 (2m_A)^2 \int\frac{d^3\mathbf k}{(2\pi)^3}
\mathcal F_A(\omega_1^2)\,\mathcal F_A(\omega_2^2)\,K\,
\mathcal C_L\,, \\
K &=& \frac{1}{2\,\omega_1\omega_2}\,
\frac{\omega_1+\omega_2}{(\omega_1+\omega_2)^2-(q^0)^2} \,, \quad \mathcal C_L = T_{00} \,,
\end{eqnarray}
in terms of the photon energies $\omega_{1,2}\simeq |\mathbf k_{1,2}|$ and 3-momenta $\mathbf k_{1,2}\equiv \mathbf q/2\pm\mathbf k$. 

The elastic hadron scattering in QCD is dominated at high energies by the Pomeron $\mathbb{P}$ exchange, 
which in the leading-log picture of perturbative QCD can be viewed as a color-singlet gluonic ladder, while 
at lower energies additional Reggeon (light $\psi\bar \psi$) exchanges contribute. In the considered dark 
Yang-Mills sector, there are no light fermions (i.e.~$m_\psi\gtrsim 5.5\Lambda_D$), so the analogous vacuum 
exchange is a purely gluonic ``dark Pomeron'' $\mathbb{P}_D$ even at low energies (zigzag line in
Fig.~\ref{fig:elastic}), which controls the forward regime of small momentum transfers, 
$|t|\lesssim \Lambda_D^2$; by contrast, $t$-channel exchange of the heavy portal fermions 
is suppressed.

In this forward limit, we evaluate the doubly-virtual Compton tensor $T^\chi_{\mu\nu}$ for $\gamma^\ast\chi\to\gamma^\ast\chi$ in the Ewerz--Maniatis--Nachtmann (EMN) framework, modeling it by exchange of a soft $C=+1$ spin-2 dark Pomeron $\mathbb{P}_D$.
This parallels the EMN description of Compton scattering off pions ($J=0$) and $\rho$ mesons ($J=1$) in QCD~\cite{Ewerz:2013kda,Britzger:2019lvc,Lebiedowicz:2022xgi}.
In the EMN model, $T^\chi_{\mu\nu}$ admits a factorized form in terms of a $\gamma\gamma\mathbb{P}_D$ vertex, the propagator $\Delta$, and a $\chi\chi\mathbb{P}_D$ vertex, supplemented by an overall elastic slope factor. The $\mathbb{P}_D$ propagator carries two symmetric index pairs and factorizes into the spin-2 projector $\mathcal P_{\kappa\lambda}^{\alpha\beta}$ and a scalar Regge factor $\mathcal R_D(W^2,t)$:
\begin{align}
\Delta_{\kappa\lambda}^{\alpha\beta} &= \mathcal P_{\kappa\lambda}^{\alpha\beta}\,\mathcal R_D=\mathcal{P}_{\kappa\lambda}^{\alpha\beta}\,
\frac{(-iW^2\alpha_D')^{\varepsilon_D + \alpha_D' t}}{W^2}\,, 
\label{eq:propagator}\\[-1pt]
\mathcal{P}_{\kappa\lambda}^{\alpha\beta} &= \tfrac{1}{4}\,\big(
g_{\kappa}^{\alpha}g_{\lambda}^{\beta}+g_{\kappa}^{\beta}g_{\lambda}^{\alpha}-\tfrac{1}{2}g_{\kappa\lambda}g^{\alpha\beta}\big)\,,~~ W^2 = (p + k_1)^2\,.
\nonumber
\end{align}
with $\varepsilon_D\,[{\rm GeV^0}]$ and $\alpha_D'\,[{\rm GeV^{-2}}]$ the intercept parameters of the soft ``dark Pomeron'' trajectory that can be matched to their QCD counterparts $\varepsilon$, $\alpha'$ known from fits to soft (low-$Q^2$) QCD data (see Table \ref{Pomeron_Parameters}). 

The effective $\gamma\gamma\mathbb{P}_D$ vertex can be written as
\begin{equation}
i\,\Gamma^{(\gamma\gamma\mathbb{P}_D)}_{\mu\nu\kappa\lambda}=
2a^\gamma_D\,\Gamma^{(0)}_{\mu\nu\kappa\lambda}-b^\gamma_D\,\Gamma^{(2)}_{\mu\nu\kappa\lambda}\,,
\label{eq:ggPvertex}
\end{equation}
where $\Gamma^{(0)}$ and $\Gamma^{(2)}$ are the standard gauge-invariant EMN tensor structures
(for on-shell photons they correspond to helicity-0 and helicity-2, respectively)~\cite{Ewerz:2013kda,Britzger:2019lvc,Lebiedowicz:2022xgi},
with $a^\gamma_D$ and $b^\gamma_D$ carrying mass dimensions $[a^\gamma_D]={\rm GeV^{-3}}$ and $[b^\gamma_D]={\rm GeV^{-1}}$.
Gauge invariance implies the Ward identities
$k_1^\mu\Gamma^{(\gamma\gamma\mathbb{P}_D)}_{\mu\nu\kappa\lambda}
=k_2^\nu\Gamma^{(\gamma\gamma\mathbb{P}_D)}_{\mu\nu\kappa\lambda}=0$.
The effective $\chi\chi\mathbb{P}_D$ vertex $\Gamma^{(\chi\chi\mathbb{P}_D)}$ has different structure for $\chi_{J=0}\equiv S$ and $\chi_{J=1}\equiv V$ dark glueballs, namely,
\begin{eqnarray}
i\Gamma^{(SS\mathbb{P}_D)}_{\alpha\beta} \!&=&\! -2i\beta^{S}_D
\Big[ (p'\!+\!p)_\alpha (p'\!+\!p)_\beta \! - \! \tfrac{1}{4} g_{\alpha\beta}(p'\!+\!p)^2 \Big]\,, \nonumber  \\[-2pt]
i\Gamma^{(VV\mathbb{P}_D)}_{\alpha\beta\rho\sigma} &=&
2a^V_D\Gamma^{(0)}_{\alpha\beta\rho\sigma}-b^V_D\Gamma^{(2)}_{\alpha\beta\rho\sigma}\,,
\label{eq:Pchichi}
\end{eqnarray}
where $\Gamma^{(0,2)}$ denote the same gauge-invariant tensor basis with the obvious replacement $(\mu,\nu,k_1,k_2)\to(\rho,\sigma,p,p')$, while $\beta^S_D\,[{\rm GeV^{-1}}]$ and $a^V_D\,[{\rm GeV^{-3}}],\, b^V_D\,[{\rm GeV^{-1}}]$ are the dark Pomeron couplings to the $S$- and $V$-glueballs, respectively. The final ingredient of the amplitude is the elastic form factor, $F_{el}=\exp[-b^D_{\rm eff}|t|/2]$, with the effective slope parameter $b^D_{\rm eff}\,[{\rm GeV^{-2}}]$ absorbing the exponential slopes of $\chi\chi\mathbb{P}_D$ and $\gamma\gamma\mathbb{P}_D$ vertices. Combining the above elements, the resulting Compton subprocess amplitudes for $S$- and $V$-glueballs are found as
\begin{align}
&T^S_{\mu\nu}= 
F_{el}\;\Gamma^{(\gamma\gamma\mathbb{P}_D)}_{\mu\nu\kappa\lambda}\;
\Delta^{\kappa\lambda\alpha\beta}\;
\Gamma^{(SS\mathbb{P}_D)}_{\alpha\beta}\,.
\label{eq:subamp-S} \\[-2pt]
&T^V_{\mu\nu,\lambda\lambda'}= 
F_{el}\;\Gamma^{(\gamma\gamma\mathbb{P}_D)}_{\mu\nu\kappa\lambda}\;
\Delta^{\kappa\lambda\alpha\beta}\;
\Gamma^{(VV\mathbb{P}_D)}_{\alpha\beta\rho\sigma}\varepsilon^{\rho}_{\lambda}\varepsilon^{\sigma}_{\lambda'}  \,,
\label{eq:subamp-V}    
\end{align}
where $\varepsilon_\lambda^\rho(p)$ and $\varepsilon_{\lambda'}^\sigma(p')$ ($\lambda,\lambda'=0,\pm1$) are the polarization vectors of the initial and final $V$-glueball states.

For concreteness, we focus on dark ${\rm SU}(3)$ and rescale key QCD inputs (e.g.~the constituent-quark mass $m_q\simeq \Lambda$ and the QCD 
confinement scale $\Lambda$) to their dark-sector counterparts. Using naive dimensional analysis, we then map 
the Pomeron intercept, effective couplings, and elastic-slope parameter extracted from Deep Inelastic Scattering 
(DIS) fits~\cite{Britzger:2019lvc} onto the corresponding dark-Pomeron parameters summarized 
in Table~\ref{tab:glueball-pars}.
%%%%%%%%%%%%%%%%%%%%%%%%%%%%%%%%%%%%%%%%%%%%%%%%%%%%%%%%%%
\begin{table}[h]
\centering
 \begin{tabular}{||c||c|c|c|c|c|c|c|c|} 
 \hline
SU(3)$_D$ & $\epsilon_D$ & $\alpha_D'$ & $\beta^{S}_D$ & $b^D_{\rm eff}$ & $a^V_D$ & $b^V_D$ & $a^\gamma_D$ & $b^\gamma_D$      \\ [1ex] 
 \hline
QCD       & $0.09$      & $0.25$      & $1.76$        & $5.0$           & $2.34$  & $4.22$   & 0.08  & 0.4  \\ [1ex] 
 \hline
$\frac{\mathrm{SU(3)}_D}{\mathrm{QCD}}$ &   $1.0$   & $\frac{\Lambda^2}{\Lambda_D^2}$  & $\frac{\Lambda}{\Lambda_D}$ & $\frac{\Lambda^2}{\Lambda_D^2}$ &  $\frac{0.6\Lambda}{m_V^2\Lambda_D}$ & $\frac{\Lambda}{\Lambda_D}$ &   
$\frac{m_q^4\alpha_D^\psi\Lambda_D}{m_\psi^4\alpha_s^*\Lambda}$ & 
$\frac{m_q^4\alpha_D^\psi\Lambda_D^3}{m_\psi^4\alpha_s^*\Lambda^3}$ 
\\ [1ex] 
 \hline

 \end{tabular}
 \caption{\label{tab:glueball-pars}
The QCD Pomeron parameters and their rescaling factors to that of the dark Pomeron $\mathbb{P}_D$. 
In the case of QCD, we take the $\mathbb{P}SS$ coupling equal to $\beta_{\pi\pi \mathbb{P}}$ 
of the pion-Pomeron $\pi\pi \mathbb{P}$ vertex and $\mathbb{P}VV$ couplings equal to those of 
$\rho$-meson -- $a_{\rho\rho\mathbb{P}}$ and $b_{\rho\rho\mathbb{P}}$. The latter satisfy 
the relation $2m_\rho^2 a_{\rho\rho\mathbb{P}} + b_{\rho\rho\mathbb{P}} = 4\beta_{\pi\pi \mathbb{P}}$ 
and have been estimated using the ratio of the corresponding $f_2$ (or $a_2$) meson couplings 
to $\rho$ found in Ref.~\cite{Ewerz:2013kda}, $b_{\rho\rho f_2}/a_{\rho\rho f_2}\simeq 3 m_\rho^2$. 
In practical calculations, we take $\alpha_s^*\equiv \alpha_s(\Lambda)\simeq 0.5$ and 
$\alpha_D^\psi\equiv \alpha_D(m_\psi)\simeq 0.12$.}
\label{Pomeron_Parameters}
\end{table}

\begin{figure}[tb]
    \vspace{0.cm}
    \includegraphics[width=0.9\linewidth]{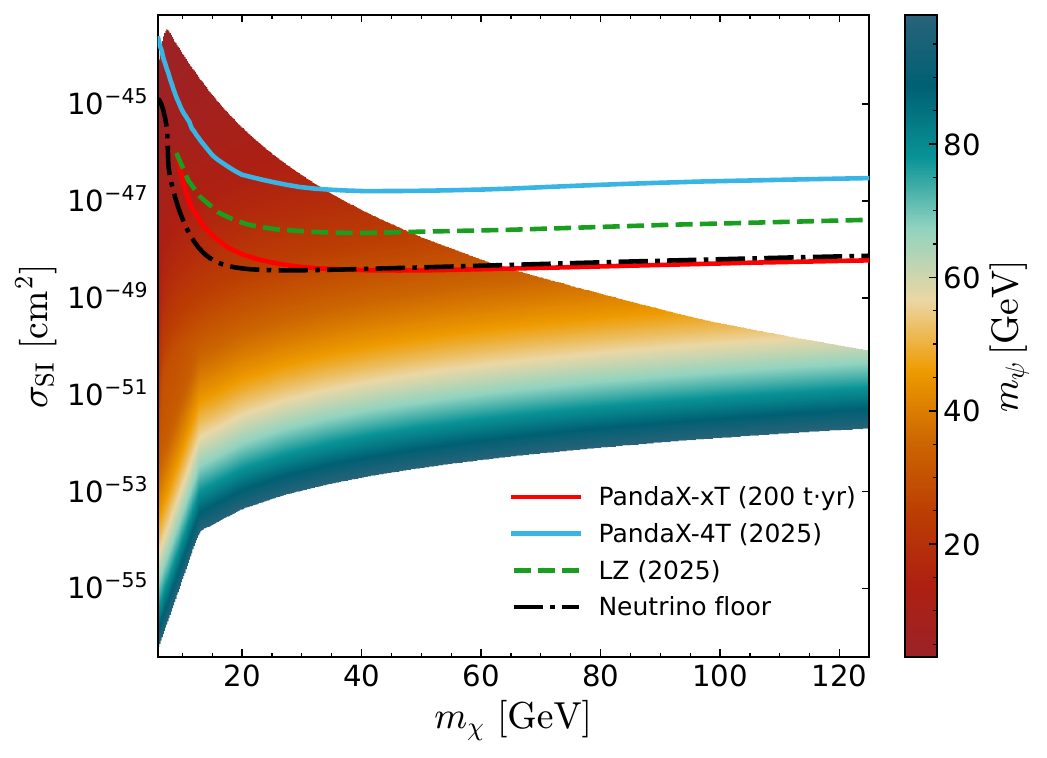}
   \caption{Spin-independent (SI) per-nucleon cross section for glueball DM 
   in the Coulomb regime compared to bounds from Xenon experiments. The colored domain 
   spans over the portal-fermion mass $m_\psi$, in the color bar, and the oddball 
   mass $m_\chi$ satisfying the glueball stability $m_\chi < 2m_\psi$. 
  Projected detector sensitivities are given for PandaX-xT~\cite{PANDA-X:2024dlo}, 
  PandaX-4T and LZ~\cite{PandaX:2024qfu,LZ:2024zvo} 
  with the neutrino floor~\cite{OHare:2021utq}.
}
    \label{fig:elastic_direct_detection}
\end{figure}

\runinsec{IV: Direct detection cross section.}{sec:Direct detection}
Starting from the Coulomb two-photon amplitude in Eq.~(\ref{eq:M-final-exactProj}), 
the elastic differential spin-independent (SI) scattering cross section off 
a nucleus $A$ in nonrelativistic regime is found as
\begin{equation}
\frac{d\sigma_{\rm el}}{dt}=\frac{|\mathcal M_\chi(s,t)|^2}{16\pi\,\lambda(s,m_A^2,m_\chi^2)},\quad
\lambda \xrightarrow{\rm NR} 4m_A^2 m_\chi^2 v_{\rm rel}^2,\,
\end{equation}
Here, $s=(p+P)^2$, $t\simeq-\bq^{\,2}=-2m_A T_A$ with $T_A\ll m_{A,\chi}$ the nuclear recoil energy, 
and $\lambda(x,y,z)=x^2+y^2+z^2-2xy-2xz-2yz$ the K\"all\'en function. To compare with experimental 
limits we quote the conventional per-nucleon normalization,
\begin{equation}
\sigma_{\rm SI}\equiv \frac{\mu_{\chi p}^2}{\mu_{\chi A}^2}\,\frac{1}{A^2}\,
\int_{t_{\rm min}}^{0}\!dt\,\frac{d\sigma_{\rm el}}{dt},
\quad t_{\rm min}\simeq-4\mu_{\chi A}^2 v_{\rm rel}^2,
\end{equation}
with $\mu_{\chi A}$ and $\mu_{\chi p}$ the DM-nucleus and DM-proton reduced masses.
Evaluating $\mathcal M_\chi$ in Eq.~(\ref{eq:M-final-exactProj}), we obtain the predicted 
$\sigma_{\rm SI}(m_\chi)$ shown in Fig.~\ref{fig:elastic_direct_detection} for different portal
mass scale $m_\psi$ in the color bar, together with the current and projected bounds 
from Xenon-based experiments. As was discussed before, the applicability of the glueball EFT 
in our analysis is justified in the heavy portal mass limit $m_\psi\gg \Lambda_D$ while 
in practice it suffices to consider $m_\psi > 5.5 \Lambda_D$ following from 
the glueball stability condition. Numerically, we find a steep parametric scaling 
$\sigma_{\rm SI}\propto \Lambda_D^{\,2.15}\,m_\psi^{-8}$. To be within the current PandaX 
sensitivity, for instance, one needs a sufficiently light portal, $m_\psi\lesssim 20~{\rm GeV}$, while PandaX-T can extend the reach up to $m_\psi\sim 30~{\rm GeV}$. Note, that the lowest $\Lambda_D^{\min}\simeq 0.55$ GeV follows from the minimal Xenon recoil energy set to $T_{\rm Xe}^{\rm min}=0.5$ keV, while the maximal recoil energy $T_{\rm Xe}^{\rm max}=2\mu_{\chi A}^2 v_{\rm rel}^2/m_A$ is set by DM kinetic energy.
%%%%%%%%%%%%%%%%%%%%%%%%%%%%%%%%%%%%%%%%%%%%%%%%%%%%
%%%%%%%%%%%%%%%%%%%%%%%%%%%%%%%%%%%%%%%%%%%%%%%%%

\runinsec{V: Electroweak completion and constraints.}{sec:UV}
The observability of glueball DM via direct-detection motivates a light charged 
dark-colored portal state with mass $m_\psi\!\lesssim\!20~{\rm GeV}$. We 
impose the inclusive LEP1 $Z$-width constraint~\cite{ALEPH:2005ab}, LEP2 
single-/multi-photon searches~\cite{Abbiendi:2000hk,Achard:2004xk}, LHC Higgs-width 
limits from coupling fits~\cite{Aad:2016neq,ATLAS:2023HinvCombo}, and oblique-parameter
bounds~\cite{PDG:2025update,Achard:2013Wpair,Peskin:1990zt,Peskin:1991sw}. 

A minimal EW completion introduces two vector-like $SU(2)_L$-singlet fermions $U,D$
with $U(1)_Y$ hypercharges $Y=\pm 1/2$ and \emph{two} vector-like $SU(2)_L$-doublet fermions $\Psi_U,\Psi_D$ with $Y=0$, while Higgs-width limits are controlled by an additional \emph{scalar}
doublet $\Phi$ (2HDM-like scenario) together with a $Z_2$ symmetry (see Appendix~\ref{app:splitmix} for more details). In a \emph{split-mixing} pattern, 
the light $\psi_\pm$ mass-eigenstates originate from different singlet-doublet blocks 
in the mass matrix. The inclusive $Z$-width bound enforces a narrow $Z$-phobic window,
\begin{eqnarray} \nonumber
    g_V^{\rm eff}(\psi_\pm) &= \mp s_W^2 \pm \frac12 f,\quad 
\Delta\Gamma_Z(Z\to \psi_\pm \bar\psi_\pm) \lesssim\ 2.3~{\rm MeV}\,,
\end{eqnarray}
where $2.3~{\rm MeV}$ is the experimental uncertainty on the total $Z$ width 
at LEP1~\cite{ALEPH:2005ab}, and $f\equiv\sin^2\theta$ is the doublet fraction 
of the light mass-eigenstates $\psi_\pm$. This implies $f\in[0.394,0.532]$ 
for $m_\psi\simeq 20~{\rm GeV}$. Moreover, in the split-mixing completion the light states $\psi_+$ and $\psi_-$ originate from
different $SU(2)_L$ doublets, so the gauge vertex $W_\mu\,\bar\psi_+\gamma^\mu\psi_-$ is absent already such that $\Delta\Gamma^{\rm tree}_W(W\to\psi_+\bar\psi_-)=0$ 
at tree level making it easier to satisfy the LEP bound, $\Delta\Gamma_W \lesssim 0.10~{\rm GeV}$~\cite{PDG:2025Wwidth,LEPEWWG:2013}, due to suppressed radiative corrections.
Besides, there are heavier Dirac partners $X^\pm$ which do not affect LEP phenomenology for $m_X \gtrsim 105~{\rm GeV}$. On the other hand, the singlet--doublet mixing required by the $Z$-phobic window breaks custodial symmetry and yields a positive fermionic shift $\Delta T_{\rm ferm}$, so $X^\pm$ cannot be arbitrarily decoupled from the EW scale. Global EW fits typically require $T\lesssim 0.2$ at 95\% CL, within the correlated $(S,T)$ ellipse~\cite{PDG:2025update,Achard:2013Wpair,Peskin:1990zt,Peskin:1991sw}. A negative scalar contribution $\Delta T_{\rm scal}<0$ from the 2HDM spectrum can reduce total $T$ well below $0.2$; see Appendix~\ref{app:splitmix}.

At LEP2 the irreducible production is $e^+e^-\rightarrow \gamma^\ast/Z^\ast\rightarrow
\psi_\pm\bar\psi_\pm$. Since the produced pair forms a confined (neutral $\rho$-meson-like) 
bound state, the visible signature is therefore governed by its subsequent 
glueball cascade and radiative $0^{++}\to \gamma\gamma$ decays in the final state. Hence, 
the LEP2 sensitivity depends on the $0^{++}$ decay length with three potential 
cases -- prompt, displaced and monophoton + missing energy ones \cite{Abbiendi:2000hk,Achard:2004xk}.
Using the scaling of dim-8 effective operator with the (lightest) portal mass, we get
\begin{equation}
c\tau_{0^{++}}\simeq 0.3~{\rm m}
\Big(\frac{m_\psi}{80~{\rm GeV}}\Big)^8
\Big(\frac{10~{\rm GeV}}{\Lambda_D}\Big)^9 \,.
\end{equation}
Our benchmark result $c\tau_{0^{++}}\simeq 5.8\, \rm{cm}$ lies in the displaced window 
which often is the hardest to constrain. It can further evade both the standard prompt and 
missing-energy searches. 

Finally, the Higgs-width constraint is implemented by requiring 
the exotic rate to obey $\Gamma(h\to\psi\bar\psi)\le \Gamma_h^{\rm SM}\,
{\rm BR}^{\rm max}_{\rm BSM}$, with ${\rm BR}^{\rm max}_{\rm BSM}\simeq 0.107$ 
from the LHC Higgs coupling fits \cite{ATLAS:2023HinvCombo,Aad:2016neq}. As shown in Appendix~\ref{app:splitmix}, this is ensured 
by \emph{sequestering} of the portal Yukawa coupling into a second Higgs doublet $\Phi$, which acquires 
the vacuum expectation value (vev) $v_\Phi$ and features a small CP-even mixing $\theta_s$ to the observed 
Higgs state. An example of a benchmark
scenario that satisfies the phenomenological bounds above is given in Table~\ref{tab:checks}.
%%%%%%%%%%%%%%%%%%%%%%%%%%%%%%%%%%%%%%%%
\begin{table}[t]
\begin{ruledtabular}
\begin{tabular}{ccccc}
$\Delta\Gamma_Z$ & $\Delta\Gamma_W$ & $c\tau_{0^{++}}$ & $T_{\rm ferm}$ & ${\rm BR}_{\rm BSM}(h)$  \\
$\lesssim 2.3~{\rm MeV}$ & $\lesssim 0.10~{\rm GeV}$ & $1~{\rm mm}$--$1~{\rm m}$ &
$\lesssim 0.2$ & $<0.107$  \\
$2.30~{\rm MeV}$ & $0$ & $5.8~{\rm cm}$ & $0.18$ & $0.088$  \\
\end{tabular}
\caption{Experimental bounds (top row) and corresponding predictions (bottom row) for our EW-complete
model benchmark point: $m_\psi = 20~{\rm GeV}$, $f = 0.394$, $m_X\simeq 119~{\rm GeV}$, 
$\Lambda_D=3.5~{\rm GeV}$, and $v_\Phi=30~{\rm GeV}$, $\theta_s=2.5\times10^{-3}$.}
\label{tab:checks}
\end{ruledtabular}
\end{table}
%%%%%%%%%%%%%%%%%%%%%%%%%%%%%%%%%%%%%%%%%%%

\runinsec{VI: Discussion and conclusions.}{sec:Conclusion}
We have shown that $C$-odd vector glueball (oddball) DM from a confining dark Yang-Mills sector can be confronted 
with direct-detection data in a controlled EFT framework. To demonstrate the power of our framework, we designed 
a minimal phenomenologically consistent formulation of light Dirac portal sector, which (i) ensures stability 
of oddball DM, and (ii) its accessibility at current and future direct-detection experiments. The major novelty
of our EFT framework is the modeling of the oddball-nucleus scattering via two-photon exchange using the dark 
tensor-Pomeron approach inspired by QCD. This has been achieved by a consistent EFT matching of the basic 
nonperturbative elements to those known from QCD phenomenology. As the main result, we find that the spin-independent 
cross section scales as $\sigma_{\rm SI}\propto \Lambda_D^{2.15} m_\psi^{-8}$ with the portal mass $m_\psi\gtrsim 5.5\Lambda_D$ and the dark confinement scale $\Lambda_D$. This result naturally delineates an essentially invisible 
multi-TeV-portal regime and a light-portal window $m_\psi\simeq 3\!-\!30$ GeV for $\Lambda_D \simeq 0.55\!-\!5.5$ GeV
potentially testable at current and next-generation Xenon-based facilities. By constructing a minimal electroweak 
completion for such a light portal sector, we demonstrate its consistency with collider phenomenology. Our results
provide a compact bridge from UV portals to nuclear recoil spectra paving a quantitative foundation for direct 
detection tests of confining dark sectors.

\emph{Acknowledgments.---} We warmly thank Jianglai Liu, Ning Zhou and Shaofeng Ge for fruitful discussions. Z.-W.W. 
is supported in part by the National Natural Science Foundation of China (Grant No.~12475105). W.W. is supported in part 
by the Natural Science Foundation of China under grants No.~12125503 and 12305103.

\appendix
\section{Electroweak Dirac portal model}
\label{app:splitmix}

\subsection{Field content and charges}
We introduce two vector-like $SU(2)_L$ singlets and two vector-like $SU(2)_L$ doublets,
\begin{align}
U &\sim (1,\;Y=+1/2),\label{eq:Ucharge_app}\\
D &\sim (1,\;Y=-1/2),\label{eq:Dcharge_app}\\
\Psi_U&=\begin{pmatrix}\psi_{U+}\\ \psi_{U-}\end{pmatrix}\sim (2,\;Y=0),\label{eq:PsiUcharge_app}\\
\Psi_D&=\begin{pmatrix}\psi_{D+}\\ \psi_{D-}\end{pmatrix}\sim (2,\;Y=0).\label{eq:PsiDcharge_app}
\end{align}
Electric charge is $Q=T_3+Y$, so $Q(U)=+1/2$, $Q(D)=-1/2$, and
$Q(\psi_{U\pm})=Q(\psi_{D\pm})=\pm 1/2$.
After EWSB, $U(1)_{\rm em}$ remains unbroken, hence the charged-fermion mass matrix is block-diagonal
in fixed-$Q$ sectors.

\subsection{Sequestered portal doublet and a $Z_2$ symmetry}
Besides the SM Higgs doublet $H$, we introduce an additional scalar doublet $\Phi$,
\begin{align}
H^T&=\left(G^+,\;\frac{v+h_0+iG^0}{\sqrt2}\right),\label{eq:Hunitary_app}\\
\Phi^T&=\left(\phi^+,\;\frac{v_\Phi+\varphi+i a}{\sqrt2}\right).\label{eq:Phiunitary_app}
\end{align}
To forbid direct portal Yukawas to $H$ while allowing Yukawas to $\Phi$, we impose a discrete symmetry in Table \ref{tab:Z2_combined}. Then $\overline{\Psi}_{U,D}\widetilde H\,U$ and $\overline{\Psi}_{U,D}H\,D$ are forbidden, but the portal Yukawas to $\Phi$
(below) are allowed. A soft $Z_2$-breaking term in the scalar potential,
\begin{align}
V(H,\Phi)\supset m_{12}^2\,H^\dagger\Phi+{\rm h.c.},
\label{eq:m12_app}
\end{align}
induces a small CP-even mixing between $h_0$ and $\varphi$, parameterized by an angle $\theta_s$.

\begin{table}[t]
\centering
\setlength{\tabcolsep}{7pt}
\renewcommand{\arraystretch}{1.15}
\begin{tabular}{c|cccccc}
\hline\hline
 & $H$ & $\Phi$ & $\Psi_U$ & $\Psi_D$ & $U$ & $D$ \\
\hline
$Z_2$ & $+$ & $-$ & $+$ & $-$ & $-$ & $+$ \\
\hline\hline
\end{tabular}
\caption{A single $Z_2$ choice that simultaneously (i) sequesters the portal Yukawa from the SM Higgs and (ii) forbids cross-sector couplings, making the split-mixing structure radiatively stable. SM fields are taken $Z_2$-even.}
\label{tab:Z2_combined}
\end{table}

\subsection{Split-mixing Yukawa sector in components}
The gauge-invariant portal Yukawas are
\begin{align}
\mathcal L_Y^{\rm portal}\supset\;
- y_U\,\overline{\Psi}_U\,\widetilde{\Phi}\,U
- y_D\,\overline{\Psi}_D\,\Phi\,D
+{\rm h.c.},
\label{eq:LYportal_app}
\end{align}
where $\widetilde{\Phi}\equiv i\sigma_2\Phi^\ast$.
Using \eqref{eq:Phiunitary_app},
\begin{align}
\widetilde{\Phi}=
\begin{pmatrix}
\dfrac{v_\Phi+\varphi-i a}{\sqrt2}\\[2pt]
-\phi^-
\end{pmatrix},
\label{eq:Phitilde_app}
\end{align}
so that \eqref{eq:LYportal_app} expands as
\begin{align}
-y_U\,\overline{\Psi}_U\widetilde{\Phi}U
&=
-\frac{y_U}{\sqrt2}(v_\Phi+\varphi-i a)\,\overline{\psi}_{U+}U
\nonumber\\[-2pt]
&\quad
+y_U\,\overline{\psi}_{U-}\phi^-U,
\label{eq:expandU_app}\\
-y_D\,\overline{\Psi}_D\Phi D
&=
-\frac{y_D}{\sqrt2}(v_\Phi+\varphi+i a)\,\overline{\psi}_{D-}D
\nonumber\\[-2pt]
&\quad
-y_D\,\overline{\psi}_{D+}\phi^+D.
\label{eq:expandD_app}
\end{align}
Only the neutral vev $v_\Phi$ generates mass mixing, and it appears \emph{only} in
$\overline{\psi}_{U+}U$ and $\overline{\psi}_{D-}D$.
Thus, in the split-mixing pattern, the components $\psi_{U-}$ and $\psi_{D+}$ do not mix at tree level.

Defining the EWSB-induced mixing masses
\begin{align}
\delta_U\equiv \frac{y_U v_\Phi}{\sqrt2},\qquad
\delta_D\equiv \frac{y_D v_\Phi}{\sqrt2},
\label{eq:deltas_app}
\end{align}
we obtain two independent $2\times2$ mixing blocks in the $Q=\pm1/2$ sectors.\\

\subsection{Charged-sector mass Lagrangian and diagonalization}
Including vector-like masses $M_U,M_D$ and singlet masses $m_U,m_D$, the charged-sector mass terms read
\begin{widetext}
\begin{align}
\mathcal L_{\rm mass}=&
-\begin{pmatrix}\bar U_L & \bar\psi_{U+}^{L}\end{pmatrix}
\bm M_{+}
\begin{pmatrix}U_R\\ \psi_{U+}^{R}\end{pmatrix}
-\begin{pmatrix}\bar D_L & \bar\psi_{D-}^{L}\end{pmatrix}
\bm M_{-}
\begin{pmatrix}D_R\\ \psi_{D-}^{R}\end{pmatrix}+{\rm h.c.}
\nonumber\\
&-M_U\,\bar\psi_{U-}\psi_{U-}-M_D\,\bar\psi_{D+}\psi_{D+},
\label{eq:Lmass_app}
\end{align}
\end{widetext}
with
\begin{align}
\bm M_{+}=
\begin{pmatrix}
m_U & \delta_U\\
\delta_U & M_U
\end{pmatrix},
\qquad
\bm M_{-}=
\begin{pmatrix}
m_D & \delta_D\\
\delta_D & M_D
\end{pmatrix}.
\label{eq:Mpm_app}
\end{align}
Each block is diagonalized by an orthogonal rotation. Defining $c_\pm\equiv\cos\theta_\pm$ and $s_\pm\equiv\sin\theta_\pm$,
\begin{align}
\begin{pmatrix}U\\ \psi_{U+}\end{pmatrix}
&=
\begin{pmatrix}
c_+ & -s_+\\
s_+ & \ \,c_+
\end{pmatrix}
\begin{pmatrix}\psi^{+}\\ X^{+}\end{pmatrix},
\label{eq:rot_plus_app}\\
\begin{pmatrix}D\\ \psi_{D-}\end{pmatrix}
&=
\begin{pmatrix}
c_- & -s_-\\
s_- & \ \,c_-
\end{pmatrix}
\begin{pmatrix}\psi^{-}\\ X^{-}\end{pmatrix}.
\label{eq:rot_minus_app}
\end{align}
The light eigenstates are
\begin{align}
\psi_{+} = c_+\,U + s_+\,\psi_{U+},\quad
\psi_{-} = c_-\,D + s_-\,\psi_{D-}.
\label{eq:lightstates_app}
\end{align}
The mixing angles satisfy
\begin{align}
\tan 2\theta_+ = \frac{2\delta_U}{M_U-m_U},\quad
\tan 2\theta_- = \frac{2\delta_D}{M_D-m_D},
\label{eq:tan2_app}
\end{align}
and we define the doublet fractions $f_\pm\equiv s_\pm^2$ (in the symmetric benchmark $f_+=f_-\equiv f$).

The corresponding eigenvalues are
\begin{align}
m_{\psi_{+}}&=
\frac{m_U+M_U}{2}-\sqrt{\left(\frac{M_U-m_U}{2}\right)^2+\delta_U^{\,2}},
\label{eq:mpsiplus_app}\\
m_{X^{+}}&=
\frac{m_U+M_U}{2}+\sqrt{\left(\frac{M_U-m_U}{2}\right)^2+\delta_U^{\,2}},
\label{eq:mXplus_app}\\
m_{\psi_{-}}&=
\frac{m_D+M_D}{2}-\sqrt{\left(\frac{M_D-m_D}{2}\right)^2+\delta_D^{\,2}},
\label{eq:mpsiminus_app}\\
m_{X^{-}}&=
\frac{m_D+M_D}{2}+\sqrt{\left(\frac{M_D-m_D}{2}\right)^2+\delta_D^{\,2}}.
\label{eq:mXminus_app}
\end{align}

\paragraph*{Comment on the $W$-width.}
The vanishing of the light--light charged current in split mixing follows from the gauge structure:
\begin{align}
\mathcal L_W \supset \frac{g}{\sqrt2}W_\mu^+\,
\Big(\overline{\psi}_{U+}\gamma^\mu\psi_{U-}
+\overline{\psi}_{D+}\gamma^\mu\psi_{D-}\Big)+{\rm h.c.},
\label{eq:LW_app}
\end{align}
which couples only within each doublet. Since $\psi_{+}$ contains $\psi_{U+}$ while $\psi_{-}$ contains $\psi_{D-}$
[cf.\ \eqref{eq:lightstates_app}], there is no tree-level $W\psi_{+}\psi_{-}$ vertex, hence
$\Delta\Gamma^{\rm tree}_W(W\to\psi_{+}\overline{\psi}_{-})=0$ independently of whether the mixing masses $\delta_{U,D}$ originate from $H$ or from $\Phi$.

\subsection{Scalar mixing and suppressed Higgs coupling to light portals}
The CP-even fields $(h_0,\varphi)$ mix into mass eigenstates $(h,h')$ via
\begin{align}
\begin{pmatrix} h\\ h'\end{pmatrix}
=
\begin{pmatrix}
\cos\theta_s & \sin\theta_s\\
-\sin\theta_s & \cos\theta_s
\end{pmatrix}
\begin{pmatrix} h_0\\ \varphi\end{pmatrix},
\qquad |\theta_s|\ll 1,
\label{eq:scalarMix_app}
\end{align}
so $\varphi = h\,\sin\theta_s+h'\cos\theta_s$.
Because the portal Yukawas are sequestered into $\Phi$ \eqref{eq:LYportal_app}, only $\varphi$ couples directly to the portal fermions.

From \eqref{eq:expandU_app}--\eqref{eq:expandD_app}, the CP-even interaction in the gauge basis can be written as
\begin{align}
\mathcal L \supset\;&
-\frac{y_U}{\sqrt2}\,\varphi\,\overline{\psi}_{U+}U
-\frac{y_D}{\sqrt2}\,\varphi\,\overline{\psi}_{D-}D
+{\rm h.c.}
\nonumber\\
=\;&-\frac{\varphi}{v_\Phi}\Big(
\delta_U\,\overline{\psi}_{U+}U
+\delta_D\,\overline{\psi}_{D-}D
\Big)+{\rm h.c.}
\label{eq:varphiGauge_app}
\end{align}
Rotating to the fermion mass basis \eqref{eq:rot_plus_app}--\eqref{eq:rot_minus_app}, the diagonal couplings to the light eigenstates are
\begin{align}
g_{\varphi\psi_{+}\psi_{+}}=\frac{\delta_U}{v_\Phi}\sin 2\theta_+ \,,\quad
g_{\varphi\psi_{-}\psi_{-}}=\frac{\delta_D}{v_\Phi}\sin 2\theta_-.
\label{eq:gvarphi_app}
\end{align}
Finally, inserting $\varphi = h\,\sin\theta_s+\cdots$ from \eqref{eq:scalarMix_app} into \eqref{eq:varphiGauge_app} yields
\begin{align}
g_{h\psi\psi}
=\sin\theta_s\; g_{\varphi\psi_{\pm}\psi_{\pm}}
\simeq
\left(\frac{\delta_{U,D}}{v_\Phi}\sin2\theta_{\pm}\right)\sin\theta_s,
\label{eq:gh_app}
\end{align}
which explains the relation $g_{h\psi\psi}=g_{\varphi\psi\psi}\sin\theta_s$ used in the main text. Here, we utilize the symmetric benchmark suggested by mass-degeneracy of the lightest eigenstates, with $g_{h\psi\psi}\equiv g_{h\psi_{+}\psi_{+}}=g_{h\psi_{-}\psi_{-}}$, $m_U=m_D$, $M_U=M_D\equiv M$, $\delta_U=\delta_D\equiv\delta$, and $\theta_+=\theta_-\equiv\theta$. Then, the partial width (per Dirac fermion) is
\begin{eqnarray}
\Gamma(h\to\psi_{\pm}\overline{\psi}_{\pm})
&=&
N\,
\frac{g_{h\psi\psi}^{\,2}\,m_h}{8\pi}\,
\beta_h^{\,3}, \label{eq:width_app} \\
\nonumber
\end{eqnarray}
where $m_h\simeq 125$ GeV is the SM Higgs boson mass, $N$ is the number of dark colors, and
\begin{eqnarray}
\beta_h &\equiv& \sqrt{1-\frac{4m_{\psi}^2}{m_h^{2}}}\,.\nonumber
\end{eqnarray}
In this case, the exotic Higgs width into the two light charged states reads
$\Gamma_{\rm BSM}(h)\simeq 2\,\Gamma(h\to\psi_{+}\overline{\psi}_{+})$.

\subsection{Oblique parameters: scalar contribution}
\label{app:ST}

The scalar doublet $\Phi$ contains a charged state $H^\pm\equiv \phi^\pm$, a CP-odd state $A\equiv a$,
and a mostly-$\Phi$ CP-even state which we denote by $\varphi$ (mass $m_\varphi$) in the small-mixing limit
$|\theta_s|\ll 1$. We denote their masses as $(m_{H^\pm},m_A,m_\varphi)$.

For completeness we write the leading one-loop scalar contributions to the oblique parameters
(in the inert/sequestered limit) as
\begin{eqnarray}
\Delta T_{\rm scal} &=&
\frac{1}{32\pi s_W^2 m_W^2}
\Big[
F(m_{H^\pm},m_A)+F(m_{H^\pm},m_\varphi)\nonumber \\
&-&F(m_A,m_\varphi)
\Big], \label{eq:Tscal_app}
\\
\Delta S_{\rm scal} &\simeq&
\frac{1}{12\pi}\ln\!\frac{m_\varphi^2}{m_A^2}\,. \label{eq:Sscal_app}    
\end{eqnarray}
Here, $F$ is the standard finite Veltman function,
\begin{equation}
F(m_1,m_2)=m_1^2+m_2^2-\frac{2m_1^2m_2^2}{m_1^2-m_2^2}\ln\!\frac{m_1^2}{m_2^2}\,.
\label{eq:VeltmanF_app}
\end{equation}
A spectrum with $m_A<m_{H^\pm}<m_\varphi$ yields $\Delta T_{\rm scal}<0$ and can partially (or fully)
compensate the positive fermionic shift $\Delta T_{\rm ferm}$.
For fixed physical scalar masses, $\Delta S_{\rm scal}$ and $\Delta T_{\rm scal}$ are essentially independent
of the CP-even mixing angle $\theta_s$, which instead controls the Higgs-width sequestering discussed above.

\bibliographystyle{bibi}
\bibliography{biblioPRL.bib}

\end{document}